\documentclass[pra,twocolumn,tightenlines,showpacs,nofootinbib]{revtex4}
\usepackage{bm,dcolumn,amsmath,graphicx}

\newcommand{\eref}[1]{Eq.~(\ref{#1})}
\newcommand{\tref}[1]{Table~\ref{#1}}

\begin{document}

\title{Enhanced sensitivity to the  fine-structure
constant variation in {Th~IV} atomic clock transition}

\author{V. V. Flambaum$^{1,2}$}
\author{S. G. Porsev$^{1,3}$}
\affiliation{$^1$ School of Physics, University of New South Wales,
                  Sydney, NSW 2052, Australia}
\affiliation{$^2$ New Zealand Institute for Advanced Study,
Massey University (Albany Campus), Private Bag 102904, North Shore MSC
 Auckland, New Zealand }
\affiliation{$^3$ Petersburg Nuclear Physics Institute, Gatchina,
                  Leningrad district, 188300, Russia}

\date{ \today }

\begin{abstract}
Our calculations have shown that the $5f_{5/2}-7s_{1/2}$ 23131 cm$^{-1}$
transition from the ground state in the ion Th$^{3+}$ is very sensitive
to the temporal variation of the fine structure constant
$\alpha=e^2/\hbar c$ ($q=-75300$ cm$^{-1}$).
The line is very narrow, the ion has been trapped and laser cooled
 and the  positive shifter line
$5f_{5/2}-5f_{7/2}$ 4325 cm$^{-1}$ ($q=+2900$ cm$^{-1}$)
may be used as a reference. A comparison may also be made with a positive shifter
in another atom or ion. This makes  Th$^{3+}$ a good candidate to search
for the $\alpha$ variation.
\end{abstract}

\maketitle

\section{Introduction}
\label{sec_intro}
Theories unifying gravity with other interactions suggest temporal and
spatial variation of the fundamental ``constants'' in expanding Universe
(see e.g. review \cite{Uza03}).
The spatial variation can explain  fine tuning of the fundamental
constants which allows humans (and any life) to appear. We appeared in
the area of the Universe where the values  of the fundamental constants
are consistent with our existence. The fundamental constants may be slightly
different near massive bodies (see e.g. review~\cite{FlaShu07}).
There are some hints for  the variation of  different fundamental constants in quasar
absorption spectra
\cite{WebFlaChu99,WebMurFla01,MurWebFla03,MurWebFla07,MurWebFla08,ReiBunHol06}
and Big Bang nucleosynthesis~\cite{DmiFlaWeb04,FlaWir07} data.
However, a majority of publications report limits on the variations
of the fundamental constants (see e.g. reviews \cite{Fla07,Lea07}).

The dependence of atomic transition frequencies on $\alpha$
may be presented in the following form
\begin{align}
\omega = \omega_\mathrm{lab} + q x, \quad x \equiv
\left({\alpha}/{\alpha_\mathrm{lab}}\right)^2 - 1\,.
\label{qfactor1}
\end{align}
In~\cite{DzuFlaWeb99L,DzuFlaWeb99A}, it was proposed to use transitions
with significantly different $q$ factors  for astrophysical and laboratory
measurements of $\alpha$ variation. One can search for a variation of $\alpha$ by comparing
two frequencies of atomic transitions over a long period of time.
Following the Ref.~\cite{DzuFlaMar03} we can represent a
measured quantity $\Delta(t)$ as
\begin{equation}
\Delta(t)=\frac{d \ln{(\omega_1/\omega_2)}}{dt} =
 \left( \frac{\dot{\omega_1}}{\omega_1} -
                   \frac{\dot{\omega_2}}{\omega_2} \right) ,
\label{Delt}
\end{equation}
where $\dot{\omega} \equiv d\omega/dt$. Taking into account
\eref{qfactor1} we can rewrite \eref{Delt} as follows
\begin{equation}
\Delta(t) \approx \left( \frac{2 q_1}{\omega_1} - \frac{2 q_2}{\omega_2} \right)
             \left( \frac{\dot{\alpha}}{\alpha_{\rm lab}} \right) .
\label{Del_t}
\end{equation}
Narrow transitions with large and different $q$ values are of experimental interest.

Note that the atomic unit of energy cancels out in the ratio
of two  transition frequencies. The $\alpha$ dependence appears due to the
relativistic corrections which rapidly increase with the nuclear charge
$Z$, $\sim Z^2 \alpha^2$,
and strongly depend on the electron angular momentum. Therefore, transitions
in heavy atoms with larger electron angular momentum difference
(like $\Delta l =2$ for $s$ and $d$ orbitals) have
larger $q$-coefficients. At present the best laboratory constraint on the
temporal variation of
$\alpha$ of $\dot{\alpha}/\alpha = (-1.6 \pm 2.3) \times 10^{-17}$ yr$^{-1}$
was obtained by Rosenband {\it et al.} in Ref.~\cite{RosHumSch08} by
comparing the frequencies of the  the $^2\!S_{1/2} \rightarrow \,
^2\!D_{5/2}$ transition in $^{199}$Hg$^+$ ($q=-52200$ cm$^{-1}$) and
 $^1\!S_0 \rightarrow \, ^3\!P^o_0$ transition in $^{27}$Al$^+$
($q=146$ cm$^{-1}$).

The relativistic corrections also rapidly increase
with the effective charge $Z_{\rm eff}$ which an external electron ``sees''.
These corrections are proportional to $Z_{\rm eff}^2$. In the case of one electron
above closed shells $Z_{\rm eff} = Z_i+1$ where $Z_i$ is the ion charge.
Therefore, highly charged ions are expected to have
larger $q$. Unfortunately, the interval between the energy levels
also increases $\sim Z_{\rm eff}^2$, therefore, there is a risk to be out of the laser
range. However, the Coulomb degeneracy and configuration crossing phenomena
may help here. Indeed, in a neutral atom an electron energy level with
larger orbital angular momentum is significantly higher than a
level with lower orbital angular momentum with the same principal quantum number $n$.

For example, in the neutral Th the $5s$ electron is a core electron while the $5f$ electron
is a valence electron. Respectively, the one-electron energy of the $5f$ electron
is much higher than that of the $5s$ electron.
Moreover, it is even higher than the energy of the $7s$ electron. As seen
from the experimental spectrum of the energy levels of the neutral (four-valence)
Th~\cite{NIST}, the energy of the $6d^2 7s5f$ state is higher than the energy of the $6d^2 7s^2$
state. In the hydrogen-like Th the energy of the $5f$ state is equal
to the energy of the $5s$ state, i.e. it is significantly lower than the
energy of the $7s$ state. Therefore,
there should be an ion charge at which the level $5f$ ``crosses'' the level $7s$
(at some higher charge it  crosses $6s$, etc). This kind of an approximate
crossing (between the $5f6d$ and the $6d7s$ states) happens in Th$^{2+}$, where the
interval between them is about 5500 cm$^{-1}$ only. The interval between
the $5f$ and $7s$ states in Th$^{3+}$ is also relatively small. Finally, the lifetime
of the excited state should be large to have a narrow line.

All these requirements clearly point towards the $5f_{5/2}-7s_{1/2}$
(23131 cm$^{-1}$) transition in Th$^{3+}$. It is also very important that
the laser cooling of the $^{232}$Th$^{3+}$ ion has recently been reported
by Campbell {\it et al.} in their paper~\cite{CamSteChu09}.
This was the first time when a multiply charged ion has been laser cooled.

According to our calculations presented below the $5f_{5/2}-7s_{1/2}$ transition
is a negative shifter with $q=-75300$ cm$^{-1}$. Another narrow line in the same
ion is the positive shifter $5f_{5/2}-5f_{7/2}$ (4325 cm$^{-1}$)
with $q=+2900$ cm$^{-1}$, this line may be used as a reference.
Comparison may also be made with a positive shifter
in another element, where $q$ may exceed $30000$ cm$^{-1}$
(see the table of atomic clock transitions with $q$ coefficients
in review~\cite{FlaDzu09} and recent work~\cite{PorFlaTor09}).
\section{Method of calculation}
\label{sec_method}
To find $q$ factors we need to solve the atomic relativistic eigenvalue
problem for different values of $\alpha$ or, respectively, for different
values of $x$ from~\eref{qfactor1}. The value of $x$ was chosen to be
equal to $|x|=1/8$. This is a convenient choice that allows us
to neglect nonlinear corrections and, on the other hand,
to make calculations numerically stable.
Thus, we need to calculate atomic frequencies $\omega_\pm$ for two
values $x=\pm 1/8$. The corresponding $q$ factor is given by
\begin{align}
q = 4 (\omega_+ - \omega_-) .
\label{qfactor2}
\end{align}

Since Th$^{3+}$ is a univalent ion we have carried out calculations
of its energy levels in the frame of the Dirac-Hartree-Fock (DHF) method
combined with many-body perturbation theory (MBPT). The latter allows us
to take into account correlations between the valence electron and the
core electrons.

We start from solving the DHF equations in the $V^{N-1}$ approximation.
On the first stage the electrons of the closed core were included in a
self-consistency procedure and their orbitals were found.
After that we constructed the valence orbitals for several
low-lying states using the frozen-core DHF equations. The virtual
orbitals were determined with the help of a recurrent
procedure described in~\cite{KozPorFla96}. The one-electron basis set of the
following size was constructed:
$ 1-20s, \, 2-20p, \, 3-20d, \, 4-25f, \, 5-18g.$

At the next stage we included core-valence correlations ($\Sigma$)
into consideration and the wave functions were determined by solving the equation
\begin{equation}
H_{\mathrm{eff}}(E_n) \, | \Psi_n \rangle = E_n \, |\Psi_n \rangle \, ,
\label{H}
\end{equation}
with an effective Hamiltonian defined as
\begin{equation}
H_{\mathrm{eff}}(E) = H_{\mathrm{FC}} + \Sigma(E) \, ,
\label{Heff}
\end{equation}
where $H_{\mathrm{FC}}$ is the frozen-core DHF
Hamiltonian and the self-energy operator $\Sigma$ is the energy-dependent
correction, involving core excitations. In the following we will refer
to it as the DHF+$\Sigma$ formalism.
\section{Discussion and results}
\label{results}
In~\tref{Tab:E} we list the low-lying energy levels and compare them with
the experimental data. To stress an importance of accounting for the
core-valence correlations we present in~\tref{Tab:E}
the results obtained on the stage of pure DHF approximation and in the frame
of DHF+$\Sigma$ formalism.

As seen from~\tref{Tab:E} on the DHF stage even the order of the low-lying
levels is incorrect. For instance, the $6d_{3/2}$ state lays deeper than
the $5f_{5/2}$ state. An agreement between theoretical and experimental
energy levels is rather poor. An accounting for the core-valence correlations
($\Sigma$ corrections) recovers the correct order of the states. Besides
that the theoretical energy levels become much closer to the experimental values.
\begin{table}
\caption{The low-lying energy levels (in cm$^{-1}$) in the DHF and
the DHF+$\Sigma$ approximations are presented.
The theoretical values are compared with the experimental data.}

\label{Tab:E}

\begin{ruledtabular}
\begin{tabular}{lclcc}
\multicolumn{2}{c}{DHF} &\multicolumn{2}{c}{DHF+$\Sigma$}
                        &\multicolumn{1}{c}{Experiment\footnotemark[1]} \\
\hline
$6d_{3/2}$ &     --- &   $5f_{5/2}$\footnotemark[2]
                                    &    ---    &     ---   \\
$6d_{5/2}$ &    4225 &   $5f_{7/2}$ &    4800   &     4325  \\
$5f_{5/2}$ &    5190 &   $6d_{3/2}$ &    9003   &     9193  \\
$5f_{7/2}$ &    8617 &   $6d_{5/2}$ &   14749   &    14486  \\
$7s_{1/2}$ &   11519 &   $7s_{1/2}$ &   21371   &    23131  \\
$7p_{1/2}$ &   46702 &   $7p_{1/2}$ &   59487   &    60239  \\
$7p_{3/2}$ &   58225 &   $7p_{3/2}$ &   72690   &    73056  \\
$8s_{1/2}$ &  102595 &   $8s_{1/2}$ &  120106   &   119622  \\
$7d_{3/2}$ &  103148 &   $7d_{3/2}$ &  120844   &   119685  \\
$7d_{5/2}$ &  104763 &   $7d_{5/2}$ &  122603   &   121427  \\
$6f_{5/2}$ &  111874 &   $6f_{5/2}$ &  128763   &   127262  \\
$6f_{7/2}$ &  112316 &   $6f_{7/2}$ &  129251   &   127815  \\
$8p_{1/2}$ &  117185 &   $8p_{1/2}$ &  135165   &   134517  \\
$8p_{3/2}$ &  122194 &   $8p_{3/2}$ &  140552   &   139871  \\
$9s_{1/2}$ &  142328 &   $9s_{1/2}$ &  161485   &   160728  \\
\end{tabular}
\end{ruledtabular}
\footnotemark[1]{Reference~\cite{NIST}}; \\
\footnotemark[2]{The removal energy of the $5f_{5/2}$ state was found to
be equal to 0.9414 au on the DHF stage and 1.0578 au on the (DHF+$\Sigma$) stage.
The experimental value is 1.0588 au}.
\end{table}

To find the $q$ factors of the excited states in respect to the ground state $5f_{5/2}$
we need to carry out calculations of frequencies $\omega_\pm$ for two
values $x=\pm 1/8$. These calculations are similar to those carried out
for the laboratory value of the fine structure constant $\alpha_{\rm lab}$. For this
reason we do not discuss them in detail. We only stress again that it is important to
include the $\Sigma$ corrections for obtaining the correct values of the
frequencies $\omega_\pm$ and, respectively, the correct values of the $q$ factors.

In~\tref{Tab:q} we present the $q$ factors of the excited states listed in
\tref{Tab:E} in respect to the ground state, obtained on the DHF+$\Sigma$ stage.
As follows from~\eref{qfactor2} the accuracy of the $q$ factors corresponds to
the accuracy of calculations of $\omega_+$ and $\omega_-$. The spectrum of the
energy levels of Th$^{3+}$ is not too dense and we believe that the energy levels
obtained at $\alpha_+$ and $\alpha_-$ are found with the same accuracy as the energy
levels computed at $\alpha_{\rm lab}$. As seen from \tref{Tab:E} the latter are reproduced
within the 10\% accuracy. Correspondingly, the accuracy of the $q$ factors can also
be estimated at the level of 10\%.
\begin{table}
\caption{The values of the $q$ factors (in cm$^{-1}$) found in
the DHF+$\Sigma$ approximations are presented.}

\label{Tab:q}

\begin{ruledtabular}
\begin{tabular}{rc}
            &    $q$  \\
\hline
 $5f_{5/2}$ &     ---  \\
 $5f_{7/2}$ &     2900 \\
 $6d_{3/2}$ &   -39000 \\
 $6d_{5/2}$ &   -34300 \\
 $7s_{1/2}$ &   -75300 \\
 $7p_{1/2}$ &   -67000 \\
 $7p_{3/2}$ &   -48900 \\
 $8s_{1/2}$ &   -57500 \\
 $7d_{3/2}$ &   -50600 \\
 $7d_{5/2}$ &   -46900 \\
 $6f_{5/2}$ &   -46100 \\
 $6f_{7/2}$ &   -45500 \\
 $8p_{1/2}$ &   -57300 \\
 $8p_{3/2}$ &   -50000 \\
 $9s_{1/2}$ &   -55100 \\
\end{tabular}
\end{ruledtabular}
\end{table}

As seen from \tref{Tab:q} almost all $q$ factors are negative.
As expected, the largest $q$ factor in absolute value
was found for the $5f_{5/2}-7s_{1/2}$ transition.
The only positive $q$ factor was obtained for the $5f_{5/2}-5f_{7/2}$ transition.

The transition frequency between fine structure levels of one multiplet
$\omega_{J,J-1}$ (where $J$ is the total angular momentum)
in the first order in $(\alpha Z)^2$ is given by the well known
Land\'{e} rule: $\omega_{J,J-1} = AJ(\alpha Z)^2$. It directly
leads to $q_{J,J-1} = \omega_{J,J-1}$ (see \eref{qfactor1}).

A marked difference between $q(5f_{5/2}-5f_{7/2}) = 2900$ cm$^{-1}$ and
$\omega (5f_{5/2}-5f_{7/2}) = 4325$  cm$^{-1}$ demonstrates an importance
of the second order relativistic corrections $\sim (\alpha Z)^4$ for Th.
These corrections modify the expression for $\omega_{J,J-1}$ leading to
(see, e.g., \cite{KozPorLev08})
\begin{equation}
\omega_{J,J-1}
=AJ(\alpha Z)^2+\left(B_J-B_{J-1}\right)(\alpha Z)^4\, ,
\label{FS}
\end{equation}
where $A$ and $B_j$ are certain coefficients and $B_j$ are not
small in comparison with $A$.

Since Th is the heavy element with $Z$ = 90,  
the parameter $(\alpha Z)$ is not small for it.
In particular, $(\alpha Z)^4 \approx 0.2$ and the term $\sim (\alpha Z)^4$
gives a noticeable contribution to the $q(5f_{5/2}-5f_{7/2})$.

To conclude, we have calculated the $q$ factors for a number of excited
states in Th$^{3+}$ in respect to the ground state $5f_{5/2}$. Our calculations
showed that the $q(5f_{5/2}-7s_{1/2})$ is very large in absolute value and,
respectively, the $5f_{5/2}-7s_{1/2}$ transition is very sensitive to the temporal variation
of the fine structure constant. Since the $7s$ state is a metastable state,
this transition is convenient for an experimental laboratory search of 
$\alpha$ variation. Another transition ($5f_{5/2}-5f_{7/2}$) can be
used as a reference. 

\section{acknowledgments}
This work was supported by the Australian Research Council and Marsden grant.
The work of S.G.P was supported in part by the Russian Foundation for Basic
Research under Grants No. 07-02-00210-a and No. 08-02-00460-a.


\begin{thebibliography}{22}
\expandafter\ifx\csname natexlab\endcsname\relax\def\natexlab#1{#1}\fi
\expandafter\ifx\csname bibnamefont\endcsname\relax
  \def\bibnamefont#1{#1}\fi
\expandafter\ifx\csname bibfnamefont\endcsname\relax
  \def\bibfnamefont#1{#1}\fi
\expandafter\ifx\csname citenamefont\endcsname\relax
  \def\citenamefont#1{#1}\fi
\expandafter\ifx\csname url\endcsname\relax
  \def\url#1{\texttt{#1}}\fi
\expandafter\ifx\csname urlprefix\endcsname\relax\def\urlprefix{URL }\fi
\providecommand{\bibinfo}[2]{#2}
\providecommand{\eprint}[2][]{\url{#2}}

\bibitem[{\citenamefont{Uzan}(2003)}]{Uza03}
\bibinfo{author}{\bibfnamefont{{\rm J-P}.}~\bibnamefont{Uzan}},
  \bibinfo{journal}{Rev. Mod. Phys.} \textbf{\bibinfo{volume}{75}},
  \bibinfo{pages}{403} (\bibinfo{year}{2003}).

\bibitem[{\citenamefont{Flambaum and Shuryak}()}]{FlaShu07}
\bibinfo{author}{\bibfnamefont{V.~V.} \bibnamefont{Flambaum}} \bibnamefont{and}
  \bibinfo{author}{\bibfnamefont{E.~V.} \bibnamefont{Shuryak}},
  \bibinfo{note}{{\rm AIP} Conf. Proc. {\bf 995}, 1 (2007); e-print
  arXiv:physics/0701220}.

\bibitem[{\citenamefont{Webb et~al.}(1999)\citenamefont{Webb, Flambaum,
  Churchill, Drinkwater, and Barrow}}]{WebFlaChu99}
\bibinfo{author}{\bibfnamefont{J.~K.} \bibnamefont{Webb}},
  \bibinfo{author}{\bibfnamefont{V.~V.} \bibnamefont{Flambaum}},
  \bibinfo{author}{\bibfnamefont{C.~W.} \bibnamefont{Churchill}},
  \bibinfo{author}{\bibfnamefont{M.~J.} \bibnamefont{Drinkwater}},
  \bibnamefont{and} \bibinfo{author}{\bibfnamefont{J.~D.}
  \bibnamefont{Barrow}}, \bibinfo{journal}{Phys. Rev. Lett.}
  \textbf{\bibinfo{volume}{82}}, \bibinfo{pages}{884} (\bibinfo{year}{1999}).

\bibitem[{\citenamefont{Webb et~al.}(2001)\citenamefont{Webb, Murphy, Flambaum,
  Dzuba, Barrow, Churchill, Prochaska, and Wolfe}}]{WebMurFla01}
\bibinfo{author}{\bibfnamefont{J.~K.} \bibnamefont{Webb}},
  \bibinfo{author}{\bibfnamefont{M.~T.} \bibnamefont{Murphy}},
  \bibinfo{author}{\bibfnamefont{V.~V.} \bibnamefont{Flambaum}},
  \bibinfo{author}{\bibfnamefont{V.~A.} \bibnamefont{Dzuba}},
  \bibinfo{author}{\bibfnamefont{J.~D.} \bibnamefont{Barrow}},
  \bibinfo{author}{\bibfnamefont{C.~W.} \bibnamefont{Churchill}},
  \bibinfo{author}{\bibfnamefont{J.~X.} \bibnamefont{Prochaska}},
  \bibnamefont{and} \bibinfo{author}{\bibfnamefont{A.~M.} \bibnamefont{Wolfe}},
  \bibinfo{journal}{Phys. Rev. Lett.} \textbf{\bibinfo{volume}{87}},
  \bibinfo{pages}{091301} (\bibinfo{year}{2001}).

\bibitem[{\citenamefont{Murphy et~al.}(2003)\citenamefont{Murphy, Webb, and
  Flambaum}}]{MurWebFla03}
\bibinfo{author}{\bibfnamefont{M.~T.} \bibnamefont{Murphy}},
  \bibinfo{author}{\bibfnamefont{J.~K.} \bibnamefont{Webb}}, \bibnamefont{and}
  \bibinfo{author}{\bibfnamefont{V.~V.} \bibnamefont{Flambaum}},
  \bibinfo{journal}{Mon. Not. R. Astron. Soc.} \textbf{\bibinfo{volume}{345}},
  \bibinfo{pages}{609} (\bibinfo{year}{2003}).

\bibitem[{\citenamefont{Murphy et~al.}(2007)\citenamefont{Murphy, Webb, and
  Flambaum}}]{MurWebFla07}
\bibinfo{author}{\bibfnamefont{M.~T.} \bibnamefont{Murphy}},
  \bibinfo{author}{\bibfnamefont{J.~K.} \bibnamefont{Webb}}, \bibnamefont{and}
  \bibinfo{author}{\bibfnamefont{V.~V.} \bibnamefont{Flambaum}},
  \bibinfo{journal}{Phys. Rev. Lett.} \textbf{\bibinfo{volume}{99}},
  \bibinfo{pages}{239001} (\bibinfo{year}{2007}).

\bibitem[{\citenamefont{Murphy et~al.}(2008)\citenamefont{Murphy, Webb, and
  Flambaum}}]{MurWebFla08}
\bibinfo{author}{\bibfnamefont{M.~T.} \bibnamefont{Murphy}},
  \bibinfo{author}{\bibfnamefont{J.~K.} \bibnamefont{Webb}}, \bibnamefont{and}
  \bibinfo{author}{\bibfnamefont{V.~V.} \bibnamefont{Flambaum}},
  \bibinfo{journal}{Mon. Not. R. Astron. Soc.} \textbf{\bibinfo{volume}{384}},
  \bibinfo{pages}{1053} (\bibinfo{year}{2008}).

\bibitem[{\citenamefont{Reinhold et~al.}(2006)\citenamefont{Reinhold, Buning,
  Hollenstein, Ivanchik, Petitjean, and Ubachs}}]{ReiBunHol06}
\bibinfo{author}{\bibfnamefont{E.}~\bibnamefont{Reinhold}},
  \bibinfo{author}{\bibfnamefont{R.}~\bibnamefont{Buning}},
  \bibinfo{author}{\bibfnamefont{U.}~\bibnamefont{Hollenstein}},
  \bibinfo{author}{\bibfnamefont{A.}~\bibnamefont{Ivanchik}},
  \bibinfo{author}{\bibfnamefont{P.}~\bibnamefont{Petitjean}},
  \bibnamefont{and} \bibinfo{author}{\bibfnamefont{W.}~\bibnamefont{Ubachs}},
  \bibinfo{journal}{Phys. Rev. Lett.} \textbf{\bibinfo{volume}{96}},
  \bibinfo{pages}{151101} (\bibinfo{year}{2006}).

\bibitem[{\citenamefont{Dmitriev et~al.}(2004)\citenamefont{Dmitriev, Flambaum,
  and Webb}}]{DmiFlaWeb04}
\bibinfo{author}{\bibfnamefont{V.~F.} \bibnamefont{Dmitriev}},
  \bibinfo{author}{\bibfnamefont{V.~V.} \bibnamefont{Flambaum}},
  \bibnamefont{and} \bibinfo{author}{\bibfnamefont{J.~K.} \bibnamefont{Webb}},
  \bibinfo{journal}{Phys. Rev. D} \textbf{\bibinfo{volume}{69}},
  \bibinfo{pages}{063506} (\bibinfo{year}{2004}).

\bibitem[{\citenamefont{Flambaum and Wiringa}(2007)}]{FlaWir07}
\bibinfo{author}{\bibfnamefont{V.~V.} \bibnamefont{Flambaum}} \bibnamefont{and}
  \bibinfo{author}{\bibfnamefont{R.~B.} \bibnamefont{Wiringa}},
  \bibinfo{journal}{Phys. Rev. C} \textbf{\bibinfo{volume}{76}},
  \bibinfo{pages}{054002} (\bibinfo{year}{2007}).

\bibitem[{\citenamefont{Flambaum}(2007)}]{Fla07}
\bibinfo{author}{\bibfnamefont{V.~V.} \bibnamefont{Flambaum}},
  \bibinfo{journal}{Int. J. Mod. Phys. A} \textbf{\bibinfo{volume}{22}},
  \bibinfo{pages}{4937} (\bibinfo{year}{2007}).

\bibitem[{\citenamefont{Lea}(2007)}]{Lea07}
\bibinfo{author}{\bibfnamefont{S.~N.} \bibnamefont{Lea}},
  \bibinfo{journal}{Rep. Prog. Phys.} \textbf{\bibinfo{volume}{70}},
  \bibinfo{pages}{1473} (\bibinfo{year}{2007}).

\bibitem[{\citenamefont{Dzuba et~al.}(1999{\natexlab{a}})\citenamefont{Dzuba,
  Flambaum, and Webb}}]{DzuFlaWeb99L}
\bibinfo{author}{\bibfnamefont{V.~A.} \bibnamefont{Dzuba}},
  \bibinfo{author}{\bibfnamefont{V.~V.} \bibnamefont{Flambaum}},
  \bibnamefont{and} \bibinfo{author}{\bibfnamefont{J.~K.} \bibnamefont{Webb}},
  \bibinfo{journal}{Phys. Rev. Lett.} \textbf{\bibinfo{volume}{82}},
  \bibinfo{pages}{888} (\bibinfo{year}{1999}{\natexlab{a}}).

\bibitem[{\citenamefont{Dzuba et~al.}(1999{\natexlab{b}})\citenamefont{Dzuba,
  Flambaum, and Webb}}]{DzuFlaWeb99A}
\bibinfo{author}{\bibfnamefont{V.~A.} \bibnamefont{Dzuba}},
  \bibinfo{author}{\bibfnamefont{V.~V.} \bibnamefont{Flambaum}},
  \bibnamefont{and} \bibinfo{author}{\bibfnamefont{J.~K.} \bibnamefont{Webb}},
  \bibinfo{journal}{Phys. Rev. A} \textbf{\bibinfo{volume}{59}},
  \bibinfo{pages}{230} (\bibinfo{year}{1999}{\natexlab{b}}).

\bibitem[{\citenamefont{Dzuba et~al.}(2003)\citenamefont{Dzuba, Flambaum, and
  Marchenko}}]{DzuFlaMar03}
\bibinfo{author}{\bibfnamefont{V.~A.} \bibnamefont{Dzuba}},
  \bibinfo{author}{\bibfnamefont{V.~V.} \bibnamefont{Flambaum}},
  \bibnamefont{and} \bibinfo{author}{\bibfnamefont{M.~V.}
  \bibnamefont{Marchenko}}, \bibinfo{journal}{Phys. Rev. A}
  \textbf{\bibinfo{volume}{68}}, \bibinfo{pages}{022506}
  (\bibinfo{year}{2003}).

\bibitem[{\citenamefont{Rosenband et~al.}(2008)\citenamefont{Rosenband, Hume,
  Schmidt, Chou, Brusch, Lorini, Oskay, Drullinger, Fortier, Stalnaker
  et~al.}}]{RosHumSch08}
\bibinfo{author}{\bibfnamefont{T.}~\bibnamefont{Rosenband}},
  \bibinfo{author}{\bibfnamefont{D.~B.} \bibnamefont{Hume}},
  \bibinfo{author}{\bibfnamefont{P.~O.} \bibnamefont{Schmidt}},
  \bibinfo{author}{\bibfnamefont{C.~W.} \bibnamefont{Chou}},
  \bibinfo{author}{\bibfnamefont{A.}~\bibnamefont{Brusch}},
  \bibinfo{author}{\bibfnamefont{L.}~\bibnamefont{Lorini}},
  \bibinfo{author}{\bibfnamefont{W.~H.} \bibnamefont{Oskay}},
  \bibinfo{author}{\bibfnamefont{R.~E.} \bibnamefont{Drullinger}},
  \bibinfo{author}{\bibfnamefont{T.~M.} \bibnamefont{Fortier}},
  \bibinfo{author}{\bibfnamefont{J.~E.} \bibnamefont{Stalnaker}},
  \bibnamefont{et~al.}, \bibinfo{journal}{Science}
  \textbf{\bibinfo{volume}{319}}, \bibinfo{pages}{1808} (\bibinfo{year}{2008}).

\bibitem[{NIS()}]{NIST}
\emph{\bibinfo{title}{{\rm NIST,} {A}tomic {S}pectra {D}atabase}},
  \urlprefix\url{http://physics.nist.gov/cgi-bin/AtData/main_asd}.

\bibitem[{\citenamefont{Campbell et~al.}(2009)\citenamefont{Campbell, Steele,
  Churchill, DePalatis, Naylor, Matsukevich, Kuzmich, and
  Chapman}}]{CamSteChu09}
\bibinfo{author}{\bibfnamefont{C.~J.} \bibnamefont{Campbell}},
  \bibinfo{author}{\bibfnamefont{A.~V.} \bibnamefont{Steele}},
  \bibinfo{author}{\bibfnamefont{L.~R.} \bibnamefont{Churchill}},
  \bibinfo{author}{\bibfnamefont{M.~V.} \bibnamefont{DePalatis}},
  \bibinfo{author}{\bibfnamefont{D.~E.} \bibnamefont{Naylor}},
  \bibinfo{author}{\bibfnamefont{D.~N.} \bibnamefont{Matsukevich}},
  \bibinfo{author}{\bibfnamefont{A.}~\bibnamefont{Kuzmich}}, \bibnamefont{and}
  \bibinfo{author}{\bibfnamefont{M.~S.} \bibnamefont{Chapman}},
  \bibinfo{journal}{Phys. Rev. Lett.} \textbf{\bibinfo{volume}{102}},
  \bibinfo{pages}{233004} (\bibinfo{year}{2009}).

\bibitem[{\citenamefont{Flambaum and Dzuba}(2009)}]{FlaDzu09}
\bibinfo{author}{\bibfnamefont{V.~V.} \bibnamefont{Flambaum}} \bibnamefont{and}
  \bibinfo{author}{\bibfnamefont{V.~A.} \bibnamefont{Dzuba}},
  \bibinfo{journal}{Can. J. Phys.} \textbf{\bibinfo{volume}{87}},
  \bibinfo{pages}{25} (\bibinfo{year}{2009}), \bibinfo{note}{e-print
  arXiv:0805.0462}.

\bibitem[{\citenamefont{Porsev et~al.}(2009)\citenamefont{Porsev, Flambaum, and
  Torgerson}}]{PorFlaTor09}
\bibinfo{author}{\bibfnamefont{S.~G.} \bibnamefont{Porsev}},
  \bibinfo{author}{\bibfnamefont{V.~V.} \bibnamefont{Flambaum}},
  \bibnamefont{and} \bibinfo{author}{\bibfnamefont{J.~R.}
  \bibnamefont{Torgerson}}, \bibinfo{journal}{Phys. Rev. A}
  \textbf{\bibinfo{volume}{80}}, \bibinfo{pages}{042503}
  (\bibinfo{year}{2009}).

\bibitem[{\citenamefont{Kozlov et~al.}(1996)\citenamefont{Kozlov, Porsev, and
  Flambaum}}]{KozPorFla96}
\bibinfo{author}{\bibfnamefont{M.~G.} \bibnamefont{Kozlov}},
  \bibinfo{author}{\bibfnamefont{S.~G.} \bibnamefont{Porsev}},
  \bibnamefont{and} \bibinfo{author}{\bibfnamefont{V.~V.}
  \bibnamefont{Flambaum}}, \bibinfo{journal}{J. \ Phys. \ B}
  \textbf{\bibinfo{volume}{29}}, \bibinfo{pages}{689} (\bibinfo{year}{1996}).

\bibitem[{\citenamefont{Kozlov et~al.}(2008)\citenamefont{Kozlov, Porsev,
  Levshakov, Reimers, and Molaro}}]{KozPorLev08}
\bibinfo{author}{\bibfnamefont{M.~G.} \bibnamefont{Kozlov}},
  \bibinfo{author}{\bibfnamefont{S.~G.} \bibnamefont{Porsev}},
  \bibinfo{author}{\bibfnamefont{S.~A.} \bibnamefont{Levshakov}},
  \bibinfo{author}{\bibfnamefont{D.}~\bibnamefont{Reimers}}, \bibnamefont{and}
  \bibinfo{author}{\bibfnamefont{P.}~\bibnamefont{Molaro}},
  \bibinfo{journal}{Phys. Rev. A} \textbf{\bibinfo{volume}{77}},
  \bibinfo{pages}{032119} (\bibinfo{year}{2008}).

\end{thebibliography}

\end{document}